\documentclass[preprint,authoryear,11pt,letterpaper]{elsarticle}
\usepackage[top=0.75in, bottom=0.75in, left=0.75in, right=0.75in]{geometry}

\usepackage{marvosym}
\usepackage{amsmath,amsthm}
\usepackage{amssymb}
\usepackage{amsfonts}
\usepackage{graphicx}
\usepackage{xcolor}
\definecolor{darkblue}{rgb}{0.0,0.5,0.5}
\definecolor{blue}{rgb}{0.0,0.5,0.68}
\usepackage[colorlinks]{hyperref}
\hypersetup{colorlinks,breaklinks,linkcolor=darkblue,urlcolor=darkblue,anchorcolor=darkblue,citecolor=darkblue}
\usepackage{booktabs}
\usepackage{threeparttable}
\usepackage{multirow}
\usepackage{lscape}
\usepackage{subcaption}
\usepackage{epstopdf}
\usepackage{lineno}

\usepackage{lineno}
\journal{}
\makeatletter
\def\ps@pprintTitle{%
   \let\@oddhead\@empty
   \let\@evenhead\@empty
   \let\@oddfoot\@empty
   \let\@evenfoot\@oddfoot
}
\makeatother

\begin{document}

\begin{frontmatter}



\title{Quantifying out-of-station waiting time in oversaturated urban metro systems}
% \tnotetext[label0]{This is only an example}

\author[label1,label2]{Kangli Zhu\corref{cor2}}
\address[label1]{State Key Laboratory of Rail Traffic Control and Safety, Beijing Jiaotong University, Beijing, 100044, China}
\address[label2]{Department of Civil Engineering, McGill University, Montreal, Quebec H3A 0C3, Canada}
% \address[label2]{Address Two\fnref{label4}}

% \fntext[label3]{I also want to inform about\ldots}
% \fntext[label4]{Small city}

\ead{zhukangli@bjtu.edu.cn}

\author[label2]{Zhanhong Cheng\corref{cor2}}
\ead{zhanhong.cheng@mail.mcgill.ca}% \ead[url]{author-one-homepage.com}

\author[label1]{Jianjun Wu}
\ead{jjwu1@bjtu.edu.cn}

\author[label1]{Fuya Yuan}
\ead{16114200@bjtu.edu.cn}

\author[label2]{Lijun Sun\corref{cor1}}
\ead{lijun.sun@mcgill.ca}

\cortext[cor2]{Contributed equally.}
\cortext[cor1]{Corresponding author.}

\begin{abstract}
Metro systems in megacities such as Beijing, Shenzhen and Guangzhou are under great passenger demand pressure. During peak hours, it is common to see oversaturated conditions (i.e., passenger demand exceeds network capacity), which bring significant operational risks and safety issues. A popular control intervention is to restrict the entering rate during peak hours by setting up out-of-station queueing with crowd control barriers. The \textit{out-of-station waiting} can make up a substantial proportion of total travel time but is not well-studied in the literature. Accurate quantification of out-of-station waiting time is important to evaluating the social benefit and cost of service scheduling/optimization plans; however, out-of-station waiting time is difficult to estimate because it is not a part of smart card transactions. In this study, we propose an innovative method to estimate the out-of-station waiting time by leveraging the information from a small group of transfer passengers---those who transfer from nearby bus routes to the metro station. Based on the estimated transfer time for this small group, we first infer the out-of-station waiting time for all passengers by developing a Gaussian Process regression with a Student-$t$ likelihood and then use the estimated out-of-station waiting time to build queueing diagrams. We apply our method to the Tiantongyuan North station of Beijing metro as a case study; our results show that the maximum out-of-station waiting time can reach 15 minutes, and the maximum queue length can be over 3000 passengers. Our results suggest that out-of-station waiting can cause significant travel costs and thus should be considered in analyzing transit performance, mode choice, and social benefits. To the best of our knowledge, this paper is the first quantitative study for out-of-station waiting time.
\end{abstract}


\end{frontmatter}

%%
%\linenumbers

\section{Introduction}\label{sec:introduction}

As the backbone of transportation systems in megacities, metro systems play a critical role in meeting the increasing demand of urban mobility. For example, the Beijing metro---a network consists of 22 lines and 391 stations---has an average daily ridership of more than 10 million by the end of 2019 \citep{beijingwiki}. To better satisfy the massive passenger demand, numerous measures have been taken to maximize the operational capacity, such as reducing peak-hour headway, increasing train speed, and removing seats for more standing space. In addition to these engineering practices, recent research also shows increasing interest in developing optimization strategies for the operation of large-scale metro systems, such as designing better timetables and schedules \citep{niu2013optimizing,sun2014demand,yin2016energy}, synchronizing different lines to reduce transfer time \citep{kang2015case}, and integrating the metro network with the bus network to minimize the impact of service disruptions \citep{jin2014enhancing,jin2015optimizing}.

Despite the tremendous efforts in increasing operational capacity, some metro systems are still operated in an oversaturated condition \citep{shi2019cooperative}, which is purely due to the fact that even the optimized capacity cannot satisfy the burst of passenger demand. As a result, it is common to see overcrowded platforms with left-behind passengers who have to wait for more than one train to get on board during peak hours \citep{zhu2018inferring}. In these extreme scenarios, safety measures need to be taken to prevent overcrowdedness on the platform and operational risks and ensure the smooth operation of the system \citep{xu2019passenger}. A common flow-control measure is \textit{out-of-station} queueing---passengers are compelled to queue outside of a metro station before entering the metro station. It is reported that the out-of-station waiting time of a few metro stations in Beijing, Guangzhou, and Shenzhen can be up to more than ten minutes at peak hours \citep{Shenzhen_queue_2021, beijing_queue_2019}.

Quantifying passenger waiting time in metro systems is crucial for evaluating service quality/performance and understanding passengers' choice behavior. In terms of the economic evaluation of public transport services, waiting time is also a critical component in assessing the social benefit/cost of different planning and operation strategies. In the literature, many methods have been developed to estimate the in-station waiting time or transfer time of metro systems using individual-based smart card transactions  \citep{sun2012using,sun2012rail,sun2015integrated,zhu2018inferring,qu2020estimating}. However, despite that out-of-station waiting may cover a substantial proportion of overall travel time and the experience is much more unpleasant than waiting inside the train station (e.g., under bad weather \citep{zhang2021outdoor}), it has received little attention in the research. This is primarily due to that out-of-station waiting time cannot be inferred directly from smart card transactions, since out-of-station waiting happens before a passenger taps into a metro station. Although one can conduct field surveys to measure out-of-station waiting, the survey approach is very time-consuming and it cannot collect data continuously for long-term monitoring.

The goal of this research is to develop a data-driven method for quantifying out-of-station waiting time using smart card data. To address the aforementioned challenges, we propose an accessible and accurate method by combining the smart card data from both bus and metro systems. Our key idea is to consider those passengers who transferred from a nearby bus stop to the metro station as a proxy, whose transfer time can be estimated as the time interval from the first tapping-out on the bus to the next tapping-in at the metro station. In detail, we first identify these transfer passengers using multi-source data. Next, the time interval between the bus tap-out and the metro tap-in is used to estimate the out-of-station queueing time. To handle the noise in the data and to extend the estimation to all passengers, we assume the latent true out-of-station waiting time is a continuous function of time and estimate it with a Gaussian Process regression with a Student-$t$ likelihood. Moreover, the estimated out-of-station waiting time is used to build queueing diagrams for further analysis. We present a case study for the Tiantongyuan North station of Beijing metro. We find the maximum out-of-station waiting time is around 15 minutes, and the maximum queue length can reach 3000 passengers.

To the best of our knowledge, this is the first quantitative study for out-of-station waiting time estimation. The contribution of this paper is three-fold. First, we propose an innovative approach that uses multi-source data and Gaussian Process regression to estimate the metro out-of-station waiting time. Our method is well-founded and can be used for long-term monitoring. Second, we show by real-world data that the out-of-station waiting is a non-negligible part of the total travel time for over-saturated metro stations; more attention should be paid on this underestimated phenomenon.

The rest of the paper is organized as follows. Section~\ref{sec:literature} reviews relevant works and presents the research gap. Section~\ref{sec:background} introduces the background and the problem. Section~\ref{sec:model} presents the modeling framework of out-of-station waiting time estimation. In Section~\ref{sec:casestudy}, we present a case study of the Tiantongyuan metro stations in Beijing. Next, we discuss potential solutions for the out-of-station waiting in Section~\ref{sec:remedies}. Finally, Section~\ref{sec:conclusion} summarizes the paper and provides future research directions.

\section{Literature review}
\label{sec:literature}

Most modern metro systems adopt a fare gantry-based smart card system, which generates a continuous flow of transactions registering when and where passengers start their trips \citep{pelletier2011smart}. Given the rich information collected, smart card data has been widely used in understanding individual travel behavior and enhancing the planning and operation of metro systems \citep[e.g., ][]{niu2013optimizing, sun2014demand, jin2014enhancing, kang2015case, jin2015optimizing, yin2016energy}. In the following, we review the application of smart card data in estimating waiting time and inferring route choices in metro systems.

The waiting time of a metro system is a crucial indicator for transit service quality, and it is also a key determinant for passenger route choice behavior \citep{wardman2004public}. Many methods have been developed to estimating waiting times from smart card data. Typically, these methods decompose the time interval between tapping-in (at origin) and tapping-out (at destination) into waiting time, onboard time, and transfer time using certain regression techniques and side information (e.g., timetables of trains). In the meanwhile, these methods usually also output the route choice of each trip. For example, \citet{kusakabe2010estimation} combined smart card data with train timetables to estimate which train is boarded by each individual traveler. \citet{sun2012using} proposed a linear regression model to decompose travel time and applied this model to estimate the spatiotemporal loading profile of trains. \citet{sun2012rail} used smart card data to study travel time reliability and proposed a probabilistic mixture model to infer passenger route choice. \citet{sun2015integrated} developed a probabilistic generative model of trip time observations characterizing both the randomness of link travel time and route choice behavior. This model can be used as a passenger flow assignment framework for service planning and operation. \citet{zhao2016estimation} proposed a probabilistic model to assign each passenger to specific trains. \citet{krishnakumari2020estimation} developed a linear regression method that estimates the delay at each metro station, link, and transfer.

The waiting time and route choice in an oversaturated metro system are more complex. For instance, passengers may travel backward to an uncrowded station to find a seat and then travel forward. \citet{tirachini2016valuation} investigated this interesting backward traveling phenomenon and estimated the disutility of sitting and standing (and also level of crowdedness) in metro trains. Besides, passengers often have to wait for multiple trains to get on board. \citet{zhu2018inferring, ma2019estimation} developed data-driven methods to estimate the number of left-behind passengers in metro systems. \citet{qu2020estimating} also studied the waiting time of left-behind passengers; they found passengers' waiting time in peak hours is much longer than the metro headway. \citet{mo2020capacity} proposed a performance monitoring framework that incorporates the number of left-behind passengers.

The aforementioned studies have proposed various methods to estimate the waiting time, transfer time, or route choice from smart card data. However, as mentioned, all these methods assume that a trip starts when a passenger taps in and finishes when he/she taps out; thus, we can see that out-of-station waiting actually leaves no traces in metro smart card data. As a result, these methods cannot quantify out-of-station waiting time, which could be a substantial component of total travel time in oversaturated stations with flow-control measures. In this paper, we combine the smart card data from bus and metro systems to infer the out-of-station waiting time in the metro system. This study is closely related to the work of \citet{sun2015characterizing}, which models passenger transfer time using smart card transactions from both bus and metro services.

\section{Background}\label{sec:background}

\begin{figure}[!ht]
\begin{center}
\includegraphics[width=0.6\textwidth]{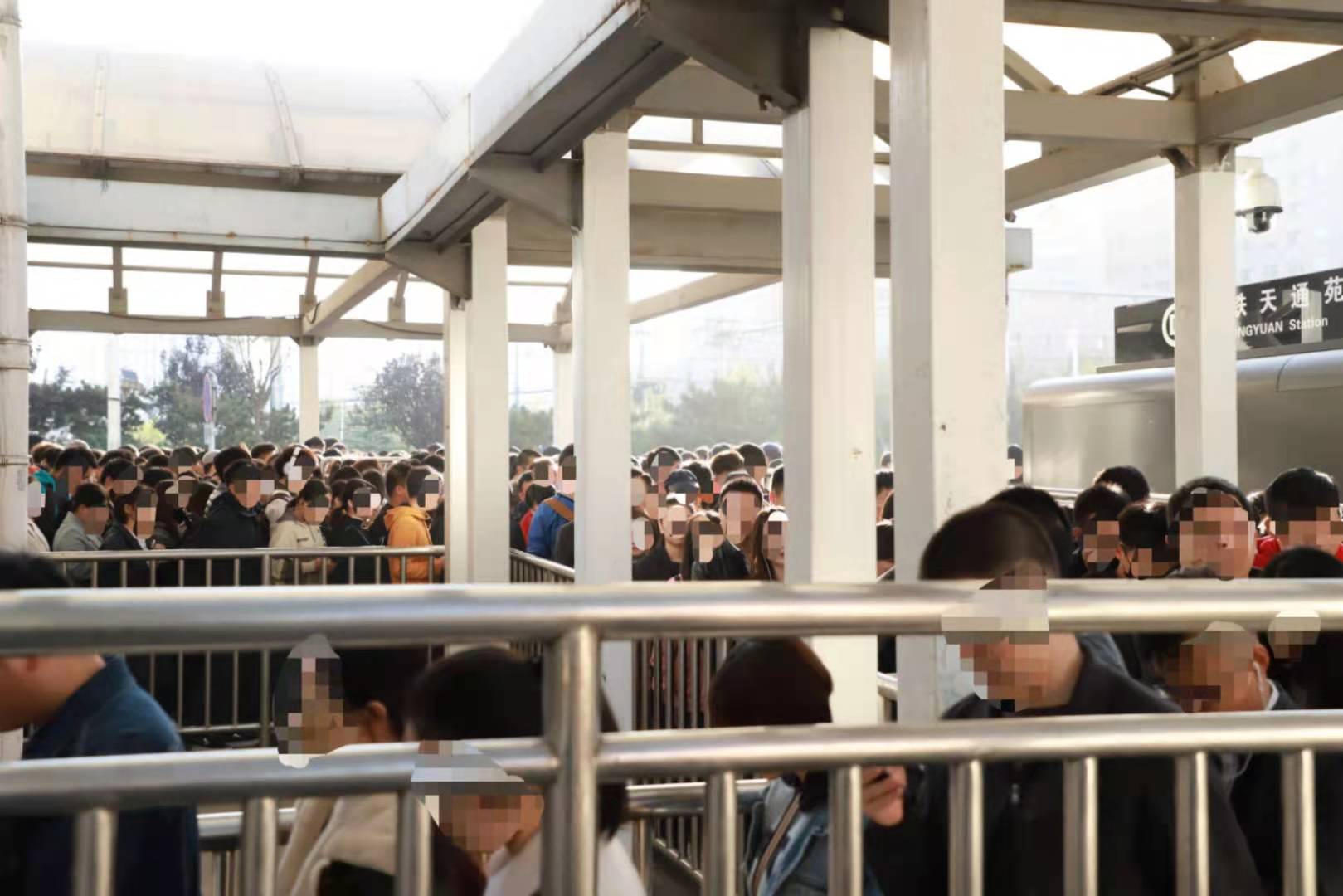}
\caption{The queue outside the Tiantongyuan metro station. Photo taken at 8:14 on Thursday, October 31, 2019.}\label{fig:waitinline}
\end{center}
\end{figure}

\begin{figure}[!ht]
\begin{center}
\includegraphics[width=0.9\textwidth]{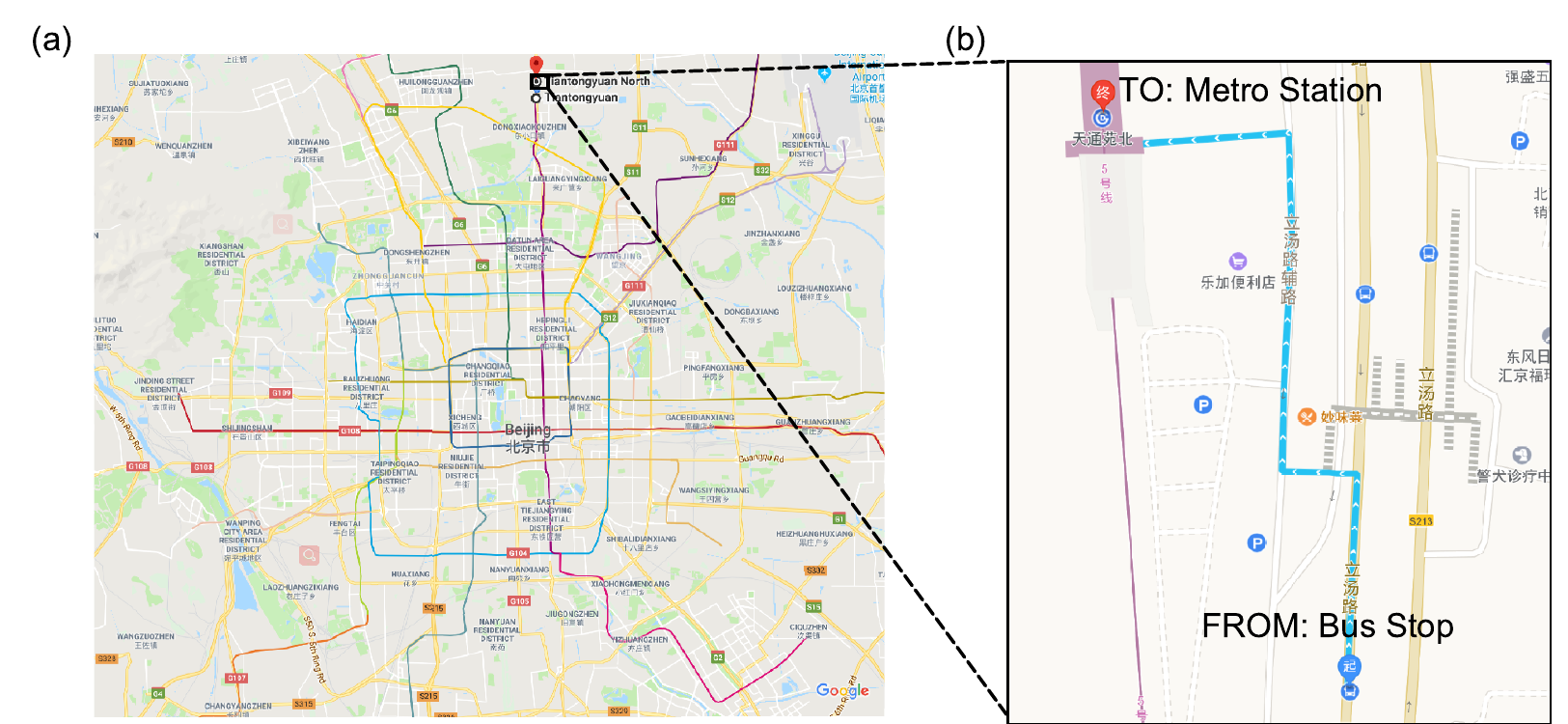}
\caption{The location of the Tiantongyuan North metro station and a nearby bus stop.}\label{fig:location}
\end{center}
\end{figure}

Beijing Metro is one of the busiest metro systems in the world. During rush hours, the ridership at a few stations is extremely high that passengers have to queue for quite a long time outside the station before entering the metro station (see Fig. ~\ref{fig:waitinline}). For example, the Tiantongyuan area of Beijing is one of the largest residential hubs in China; it has a total population of 700,000 in 2019 \citep{tiantongyuanwiki}. There are three metro stations, Tiantongyuan North (TTY-N), Tiantongyuan (TTY), and Tiantongyuan South (TTY-S), in this area. Due to the large number of commuting passengers, all three stations are oversaturated during morning peak hours on weekdays. In this paper, we use the TTY-N station as an example to demonstrate our proposed solution to quantify out-of-station waiting time. The location of the TTY-N station is shown in Fig.~\ref{fig:location}. Because the TTY-N station is the northern terminus of Metro Line 5, the boarding rate in the morning peak is controlled to alleviate the overcrowdedness on the platform and to prevent the service at downstream stations be overwhelmed; this is also one of the reasons for the out-of-station queueing. Without this flow-control intervention, the trains will be fully loaded at departure, leaving no capacity for passengers waiting at the subsequent/downstream stations.

The public transit system in Beijing uses a distance-based fare scheme. Therefore, passengers need to tap their smart cards or tickets when getting on and off a bus and when entering and leaving a metro station. Useful information in smart card data includes anonymous IDs, origin/destination, and timestamps of tapping-in/-out. For bus trips, the transactions also record the ID of the bus. Note that user IDs are consistent in both metro and bus systems, so that we can link trips from both systems for a particular user. Next we show how to estimate the out-of-station waiting time by combining the smart card data from both bus and metro systems.

\begin{figure}[!ht]
    \centering
    \includegraphics[scale=0.5]{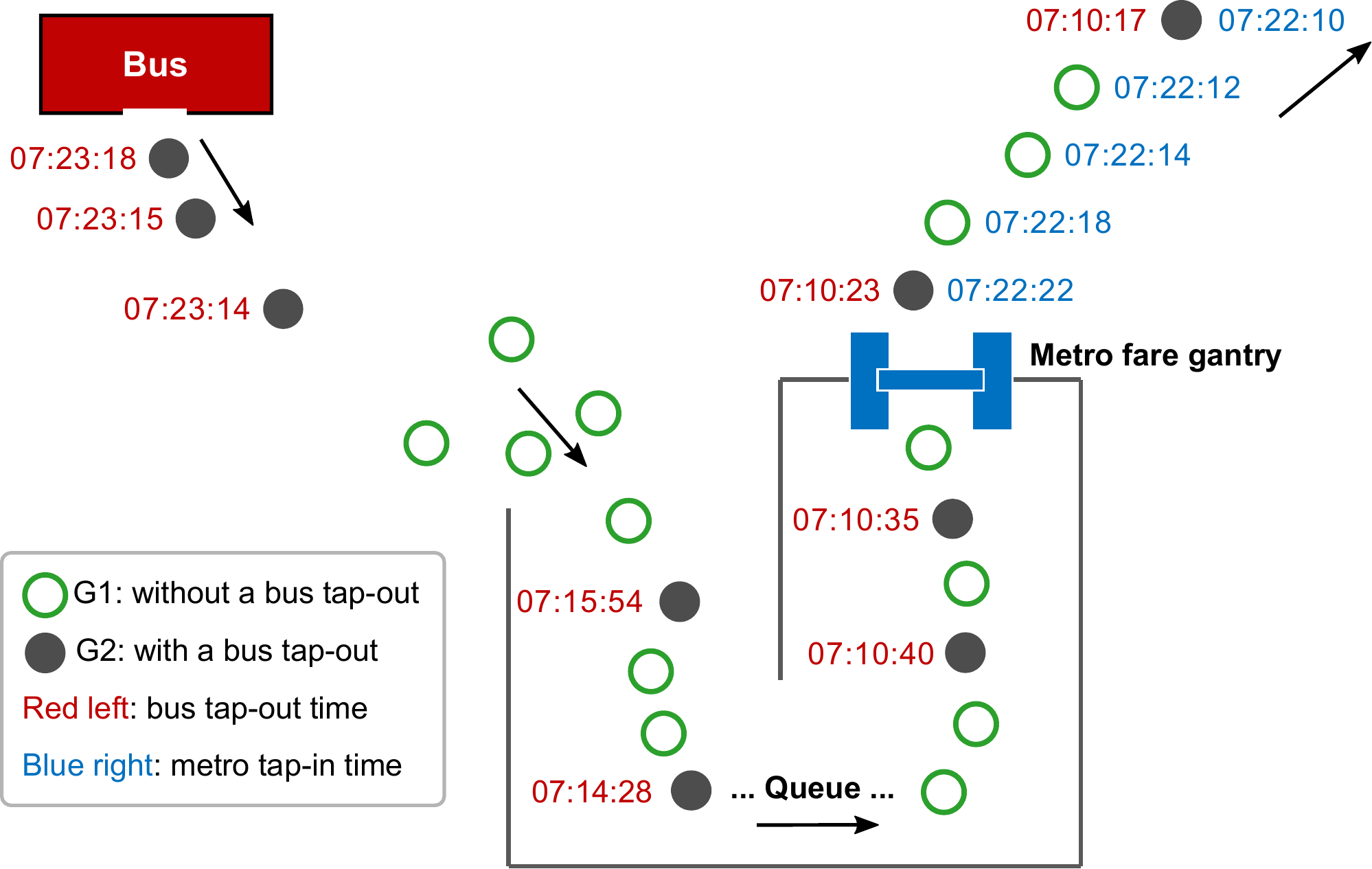}
    \caption{An illustration of out-of-station queueing of a metro station. G1 means direct passengers without a previous bus trip. G2 represents transfer passengers coming from a previous bus trip. Red numbers to the left side give the bus tap-out time $t_\text{out}$, and blue numbers to the right side give the tap-in time $t_\text{in}$ of the metro station. We estimate the out-of-station queueing time for both G1 and G2 based on the $t_\text{in}$ and $t_\text{out}$ of G2.}
    \label{fig:transfer_trip}
\end{figure}

As illustrated in Fig.~\ref{fig:transfer_trip}, we separate all the incoming passengers at a metro station into two groups: (G1) direct passengers who do not have a previous bus transfer and (G2) transfer passengers that coming from a nearby bus stop. For a direct passenger $i$, we only know the tap-in time $t_{\text{in},i}$ at the metro fare gantry, but we have no information about the out-of-station queueing. For a transfer passenger $i$, we can know the metro tap-in time $t_{\text{in},i}$, the bus tap-out time $t_{\text{out},i}$, and the transfer duration $d_{\text{transfer},i} = t_{\text{in},i} - t_{\text{out},i}$ (we use $t$ for a timestamp and $d$ for a time duration/interval). In addition, we can estimate the out-of-station waiting duration for a passenger in G2 by subtracting the walking time $d_{\text{walk},i}$ between the bus stop and the metro station from the transfer duration.

Typically, G2 only accounts for a small percentage of the total boarding passengers. Therefore, we regard G2 as a small sample set drawn from total passengers and use it to estimate the out-of-station waiting profile for all boarding passengers. In doing so, we need to (1) accurately estimate the out-of-station waiting time for all boarding passengers and (2) develop a method to analyze queueing profile. We introduce the methodologies for these tasks in Section~\ref{sec:model}.

% For a passenger in Group (2), the smart card system will create two transactions: the first transaction comes from the bus trip and second transaction comes from the follow-up metro trip. The interval from tapping-out timestamp ($t_{\text{out}}$) of the first bus transaction to the tapping-in timestamp $t_{\text{in}}$ of the second metro transaction captures all the additional cost on walking and out-of-station waiting. Ideally, a passenger on the bus will tap out when he/she alights from the bus. However, in operation it is very common for a passenger to swipe the card earlier to speed up the alighting process. To address the error caused by this, we replace tapping-out time for passengers on the same bus with that of the last passenger.

\section{Modeling framework}
\label{sec:model}
This section elaborates the methods for profiling the out-of-station waiting time using smart card data. First, in Section~\ref{sec:GP}, we illustrate the impact of noise in the data and propose a Gaussian Process regression for the out-of-station waiting time estimation. Then, in Section~\ref{sec:queue}, we introduce the idea of using a queueing diagram to analyze the out-of-station waiting.

\subsection{Gaussian Process for waiting time estimation}\label{sec:GP}

\begin{figure}[!ht]
\begin{center}
\includegraphics[]{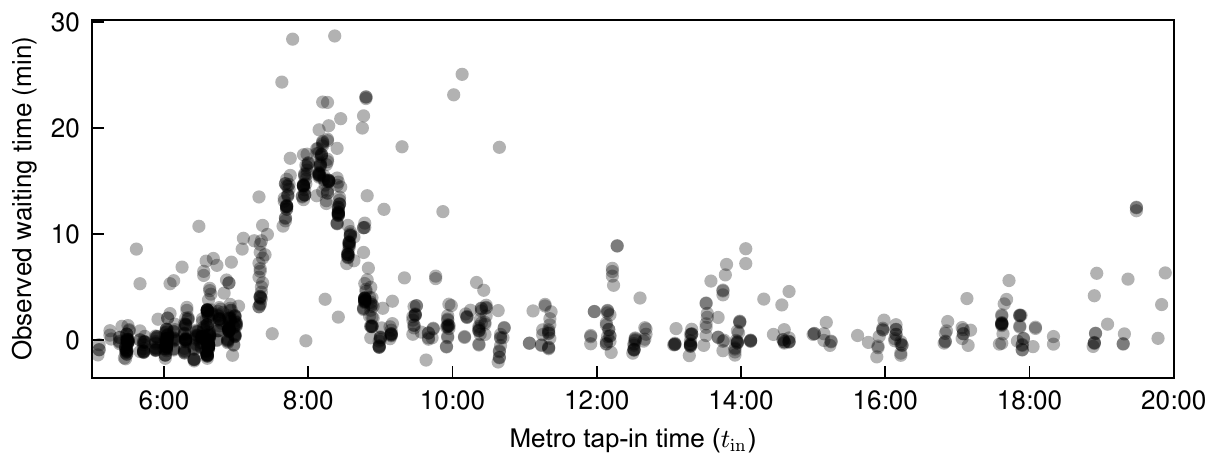}
\caption{The observed out-of-station waiting time (i.e., $d_\text{transfer} - d_\text{walk}$) at different metro tap-in times in a workday.}\label{fig:observation}
\end{center}
\end{figure}

The out-of-station waiting time of a passenger $i$ in G2 can be roughly estimated by $d_{\text{transfer},i} - d_{\text{walk},i}$, and we refer it as the \textit{observed} waiting time for simplicity. Fig.~\ref{fig:observation} shows the observed waiting time at different metro tap-in times, where we regard the walking time $d_\text{walk}$ as a constant and determine it by the median value of all $d_\text{transfer}$ during 12:00-4:00 pm (assuming no out-of-station waiting at off-peak hours). We can see the observed waiting time is much higher in the morning peak. However, there is substantial noise in the observed waiting time. Even in a short period of time, there are significant discrepancies between different observations. Sources of noise include different walking speeds, unsynchronized clocks between smart card readers, intermediate activities, and some passengers may ``tap-out'' before the bus arrives at the bus stop to speed up the alighting. Because of the noise, a well-founded method is required for out-of-station waiting time estimation.

We use a Gaussian Process (GP) regression \citep{williams2006gaussian} to estimate the out-of-station waiting time. A GP is a non-parametric Bayesian model that defines a distribution over functions. Gaussian processes are very flexible and can approximate complex functions using various kinds of kernels and likelihoods. Most importantly, a GP is a probabilistic approach that also gives confidence intervals for the estimated values. We refer readers to \citet{williams2006gaussian} for more information about Gaussian Processes.

Let $y\left(t\right)$ be the observed waiting time for a passenger with metro tap-in time $t$. For ease of description, the time $t$ in here and after means the metro tap-in time if not otherwise specified. The observed waiting time can be decomposed into a latent ``true'' waiting time $f(t)$ and a noise term $\varepsilon$:
\begin{equation}\label{eq:observation}
    y\left( t \right) =f\left(t \right)+\varepsilon.
\end{equation}
We assume the ``true'' out-of-station waiting time $f\left(t\right)$ is a continuous function of time. We need to infer the latent ``true'' waiting time given the observed waiting time. In doing so, we impose a GP prior to $f\left(t\right)$, meaning the function's values $f\left(\mathbf{t}\right)=\left[f\left(t_1\right), f\left(t_2\right),\cdots ,f\left(t_n\right)\right]^\top$ for any finite collection of inputs $\mathbf{t}=\left[t_1, t_2, \cdots, t_n \right]^\top$ have a joint multivariate Gaussian distribution. We will write
\begin{equation}\label{eq:GP}
    f\left( t \right) \sim \mathcal{GP}\left(\mu\left( t \right),  k\left(t, t^{\prime}\right) \right),
\end{equation}
where $\mu\left(t \right)$ is the mean function and $k\left(t, t^{\prime}\right)$ is the covariance/kernel function. By convention, the mean function is set to zero, i.e., $\mu\left(t\right)=0$. For the covariance function, we choose the commonly used squared-exponential kernel
\begin{equation}\label{eq:kernel}
    k\left(t, t^{\prime} \mid \ell, \lambda^2 \right)=\lambda^{2} \exp \left(-\frac{\left(t-t^{\prime}\right)^{2}}{2 \ell^{2}}\right),
\end{equation}
where the length scale $\ell$ and the variance $\lambda^2$ are two hyperparameters that should be calibrated by data. The covariance in Eq.~\eqref{eq:kernel} is larger for two closer $t$ and $t^{\prime}$, indicating passengers that enter the metro station at a closer time are more likely to have more similar waiting time. We can see a GP is fully specified by the mean and covariance functions and does not impose any assumption on the form of the function $f\left(t\right)$.

When using an i.i.d.\ Gaussian distribution for the noise term $\varepsilon$, the posterior of the latent variable $f(\mathbf{t})$ can be solved analytically. However, this convenient approach is very sensitive to outliers and is not appropriate for our data. To make a robust estimation for the ``true'' waiting time, we assume the noise is a zero-mean i.i.d.\ Student-$t$ distribution with a long-tail probability density function \citep{jylanki2011robust}:
\begin{equation}\label{eq:t}
    p\left( \varepsilon \mid \nu, \sigma \right) =\frac{\Gamma\left(\frac{\nu+1}{2}\right)}{\Gamma\left(\frac{\nu}{2}\right) \sqrt{\pi \nu }\sigma}\left(1+\frac{\varepsilon^2}{\nu\sigma^2}\right)^{-\frac{\nu+1}{2}},
\end{equation}
where $\nu$ is the degrees of freedom and $\sigma$ the scale parameter \citep{gelman2013bayesian}.

Eq.~\eqref{eq:observation}--\eqref{eq:t} describe the GP regression with a Student-t likelihood. Given a set of observed waiting times $\mathbf{y}$ at metro tap-in time $\mathbf{t}$, the four hyperparameters $\theta=\lbrace \ell, \lambda^2, \nu, \sigma\rbrace$ can be optimized by maximizing the log marginal likelihood
\begin{equation}
    \log p(\mathbf{y} | \mathbf{t}, \theta) = \log\int p(\mathbf{y} \mid \mathbf{f}) p(\mathbf{f} \mid \mathbf{t}, \theta) d\mathbf{f}.
\end{equation}
The log marginal likelihood cannot be explicitly obtained when the noise is a Student-$t$ distribution. Therefore, approximate inference methods \citep{neal1997monte, vanhatalo2009gaussian, jylanki2011robust} were developed to fit hyperparameters. We use the Laplace appropriation \citep{vanhatalo2009gaussian} as implemented in GPy \citep{gpy2014} for the appropriate inference. Next, for new passengers with metro tap-in times $\mathbf{t}_{*}$, we can calculate the posterior distribution $p\left(\mathbf{f}_{*} \mid \mathbf{y}, \mathbf{t}, \mathbf{t}_{*} \right)$ of their ``true'' out-of-station waiting time. The posterior distribution is a Gaussian but is also approximately solved \citep{jylanki2011robust}. We use the posterior mean as a point estimation $\hat{f}(\mathbf{t}_{*})$ for the out-of-station waiting time, referred to as the \textit{estimated} waiting time in the following.

% After determining the hyperparameters, we can use the of any passengers by their metro tap-in time $\mathbf{t}_{*}$.

\subsection{Queueing diagram}\label{sec:queue}
\begin{figure}[!ht]
    \centering
    \includegraphics[scale=0.8]{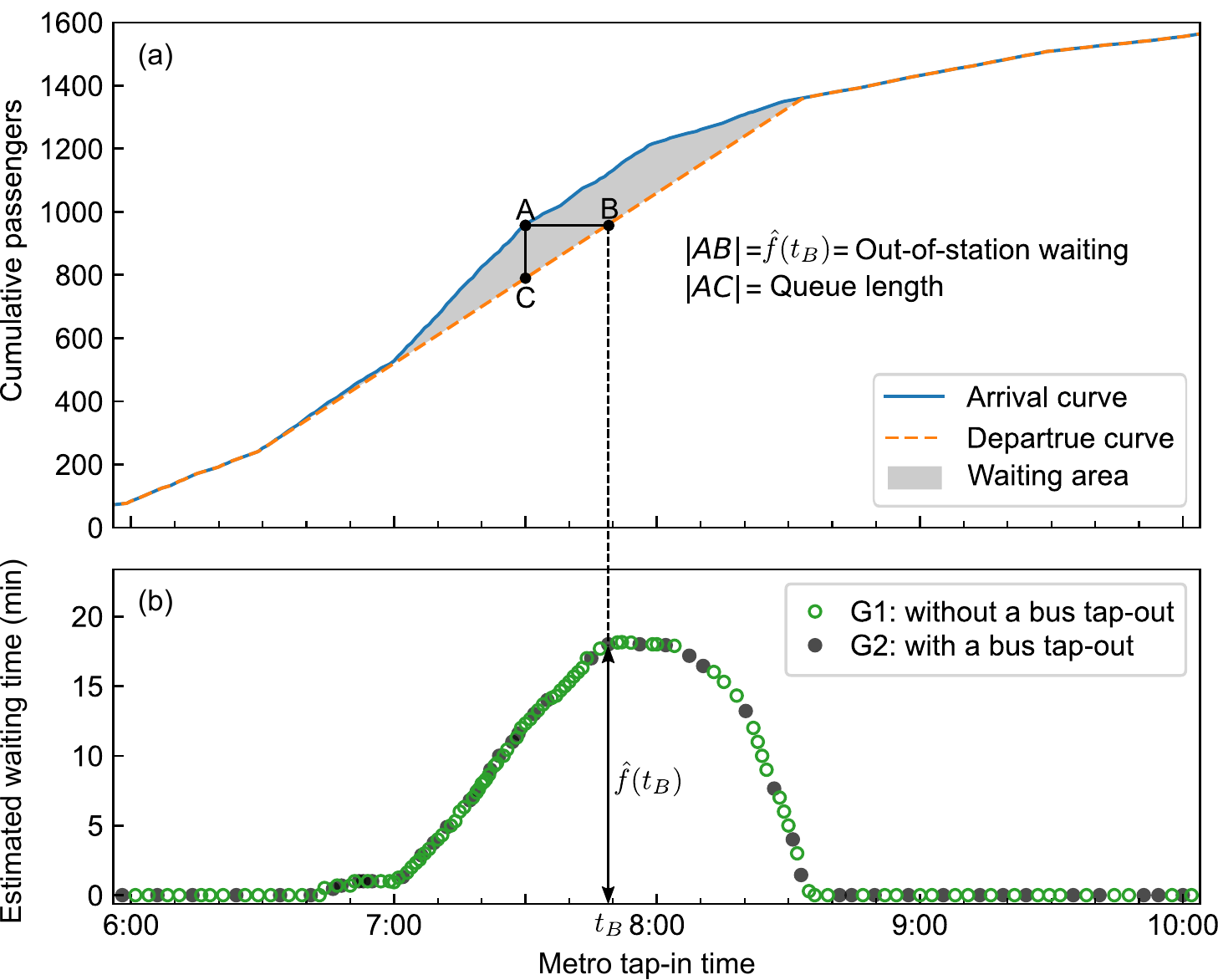}
    \caption{Establishing a queueing diagram by the estimated out-of-station waiting time -- a virtual example. (a) The queueing diagram. (b) The estimated out-of-station waiting time (only shows 10\% of passengers for clarity).}
    \label{fig:queueing}
\end{figure}

The out-of-station waiting phenomenon at a metro station is a queueing process with varying arrival rate and service rate. To better analyze the reason and the impact of the out-of-station queue, we further establish a queueing diagram based on the estimated waiting time, as illustrated in the virtual example of Fig.~\ref{fig:queueing}.

We used the virtual example in Fig.~\ref{fig:queueing} to illustrate how to establish a queueing diagram from the estimated out-of-station waiting time. Fig.~\ref{fig:queueing} (a) shows the queueing diagram, where the departure curve indicates the service rate at the metro gantries. Because the smart card data in Beijing contain passengers' tap-in times, the departure curve is directly reconstructed from passengers' metro tap-in records. The passenger arrival curve is not directly available from the data but can be inferred from the metro tap-in time and the out-of-station waiting duration. For example, the point $B$ in Fig.~\ref{fig:queueing} represents a passenger that tapped in the metro station at $t_B$; we can estimate his/her out-of-station waiting duration $\hat{f}(t_B)$ by the GP model described in Section~\ref{sec:GP}, as shown in Fig.~\ref{fig:queueing} (b). The segment $|AB|$ in the queueing diagram means the out-of-station waiting duration, we can thus calculate the arrival time for this passenger (point $A$) by $t_B - \hat{f}(t_B)$. The arrival curve can thus be obtained by connecting the estimated arrival times of all passengers. Note we use the estimated waiting time instead of the observed waiting time for both G1 and G2 to avoid the impact of noise in the data.

The queueing diagram provides important information much more than just a visualization. For example, the slope of the departure curve represents the service rate at the metro entrance, the slope of the arrival curve means the arrival rate. The horizontal distance (e.g., $|AB|$) and vertical distance (e.g., $|AC|$) between the two curves represent the waiting time and the queueing length, respectively. Moreover, the area between the two curves represents the total waiting time of all passengers. Next, we will build queueing diagrams to analyze the out-of-station queueing at the TTY-N station.

\section{Results}
\label{sec:casestudy}
In this section, we present the results for the out-of-station queueing at the TTY-N metro station. Firstly, the data and the demand pattern at the TTY-N station are introduced in Section~\ref{sec:case_demand}. Next, Section~\ref{sec:case_waiting} exhibits the out-of-station waiting time estimated by GP regressions. Finally, we analyze the queueing process in the morning peak and discuss possible solutions in Section~\ref{sec:case_queue}.

\subsection{Data description}\label{sec:case_demand}
Based on the available data, we select a five-day period from August 3rd to 8th, 2015, to analyze the out-of-station queueing at the TTY-N metro station. This five-day period reflects a typical weekday demand pattern. We have full smart card data for the TTY-N station in this period. The tap-in/out information for all metro passengers, including those who use tickets, is registered in the data. Besides, we also have smart card data from three different bus routes that pass through the TTY-N bus stop (next to the TTY-N subway station). The walking time from the bus stop to the metro station is the same for the three bus routes. The whole analysis is based on the tap-out at the TTY-N bus stop and the tap-in at the TTY-N metro station.

\begin{figure}[ht]
    \centering
    \includegraphics[scale=0.8]{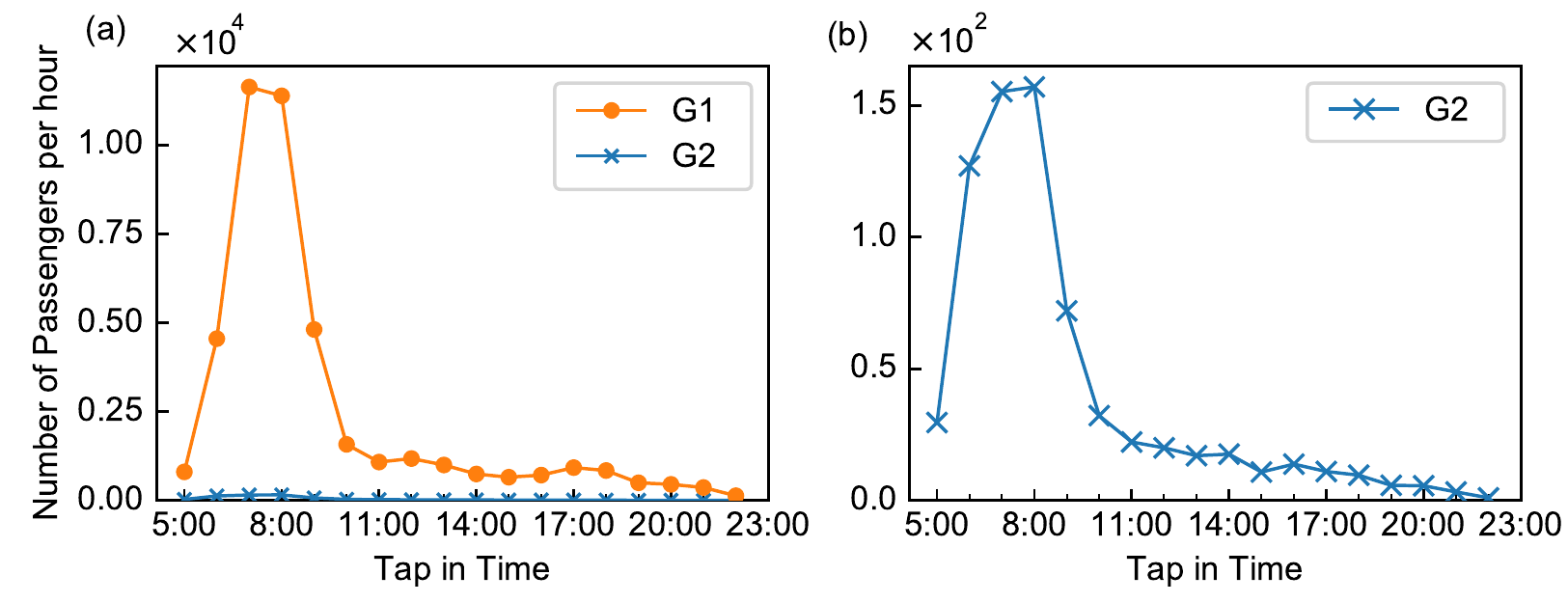}
    \caption{The number of boarding passengers per hour at the TTY-N station (Monday, August 3rd, 2015).}
    \label{fig:Subwaydemand}
\end{figure}

If we detect the same smart card ID has an immediate metro trip after a bus trip, and the interval between the bus tap-out and the metro tap-in is within 30 minutes, we regard this ID as a passenger in G2. After separating passengers into G1 and G2, the boarding demand of G1 and G2 at the TTY-N station on a typical weekday is shown in Fig.~\ref{fig:Subwaydemand}. we can see the TTY-N station services more than ten thousand passengers per hour in the morning peak (7-8 am), and the demand is very low in the afternoon and evening, showing TTY-N is a typical residential-type station. On the other hand, the number of passengers in G2 is much smaller than that in G1. This is because only a small portion of passengers are transferred from the bus stop. We use G2 as a small sample drawn from all passengers to recover the out-of-station waiting profile at the TTY-N station.

\subsection{Estimated out-of-station waiting}\label{sec:case_waiting}
\begin{figure}[!t]
    \begin{center}
    \includegraphics[]{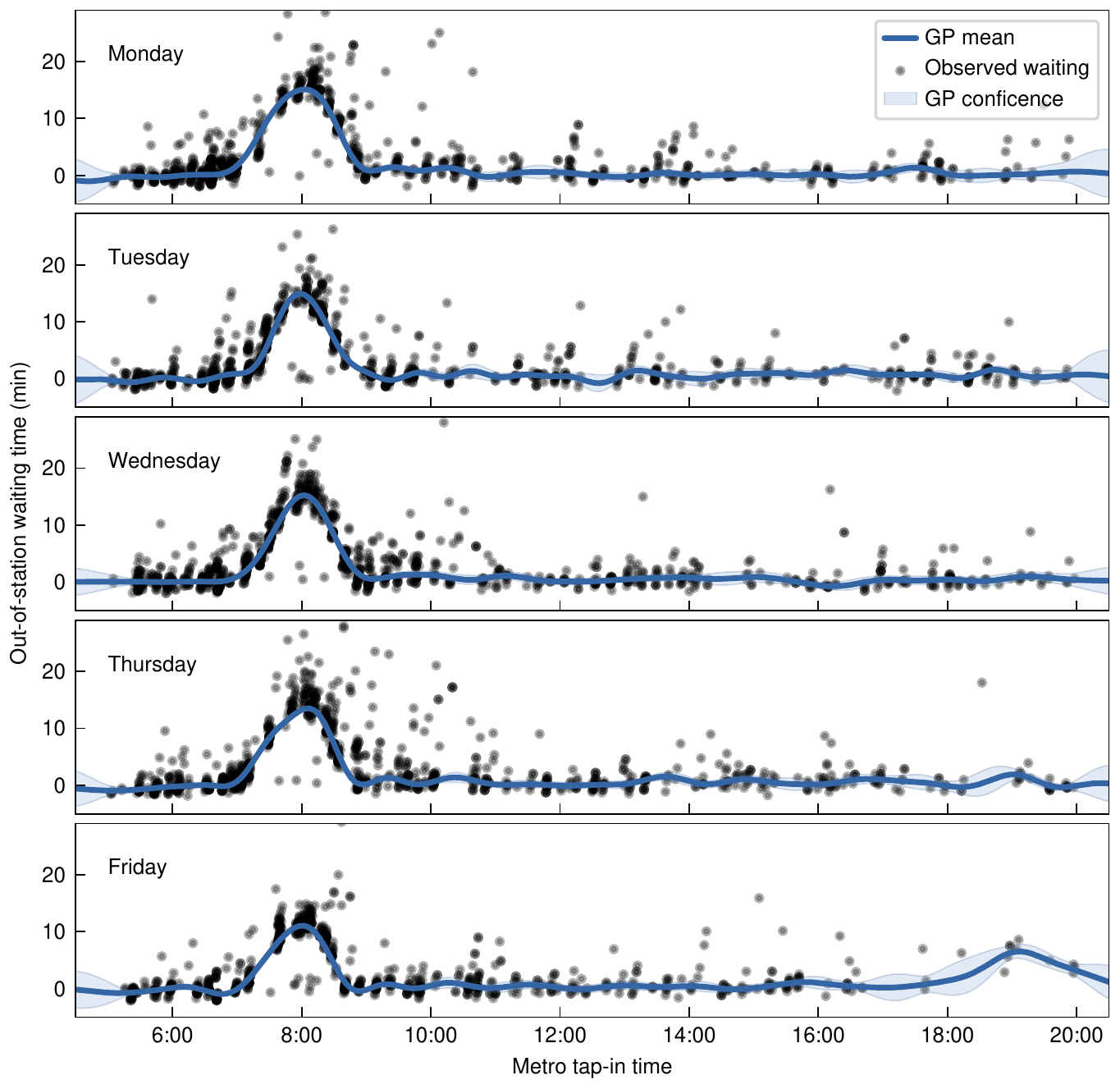}
    \caption{The observed waiting time, the GP posterior mean, and the 95\% confidence interval of the GP posterior mean.}\label{fig:GP_result}
    \end{center}
\end{figure}

Because the TTY-N station has a low boarding demand in off-peak hours, we can safely assume there is no out-of-station waiting at off-peak hours. Therefore, we regard the walking time as a constant and determine it by the median value of all $d_{\mathrm{transfer}}$ during 12:00-4:00 pm. Next, we can calculate the observed waiting time, as shown in the black points in Fig.~\ref{fig:GP_result}. We can see the noise is very high for the observed waiting time.

Next, we fit a GP for each day. Fig.~\ref{fig:GP_result} shows the posterior mean and confidence interval of the estimated out-of-station waiting time. We can see the GP with a Student-$t$ likelihood is robust to the noise, and the estimated waiting time makes intuitive sense. Despite the presence of outliers, the estimated waiting time in general wiggles around zero from 5:00 to 7:00 am and after 9:00 am, which is consistent with the real-life situation. We also see the confidence interval is larger in the period with fewer observations. The estimated waiting time on Friday night is unusually high, this could be caused by too few G2 passengers during that period. We expect the estimated waiting time for morning peaks to be more reliable because of the larger number of observations. All five weekdays have significant out-of-station waiting from around 7:00 to 9:00 am. The maximum waiting time is around 15 minutes for Monday to Thursday, and the waiting time on Friday is relatively shorter. The quantitative results for the waiting time is shown in the Tab.~\ref{tab:result}.

\subsection{Queueing analysis}\label{sec:case_queue}
\begin{figure}[!t]
    \begin{center}
    \includegraphics[]{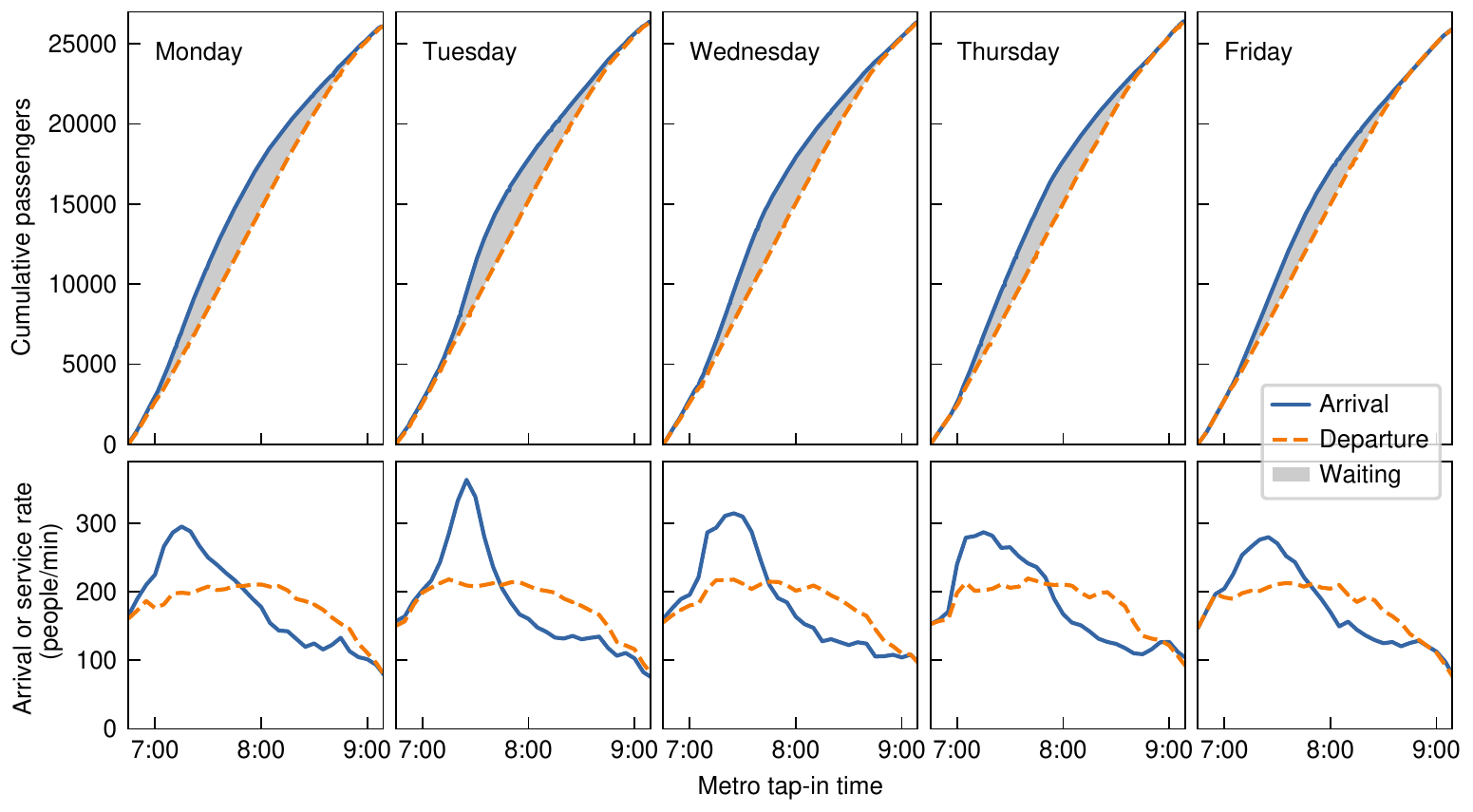}
    \caption{The queueing process in the morning peak.}\label{fig:case_queue}
    \end{center}
\end{figure}

This section analyzes the out-of-station waiting by queueing diagrams. We first set negative values in the estimated waiting time to zero. Following the illustration in Fig.~\ref{fig:queueing}, we next establish queueing diagrams for the five weekdays, as shown in Fig.~\ref{fig:case_queue}.

The upper half of Fig.~\ref{fig:case_queue} shows the cumulative arrival curve and departure curve at the TTY-N metro station. Note that the ``departure'' means entering the metro gantry rather than boarding a train. The arrival/service rate at time $t$ is estimated by the average arrival/service rate in $[t-5\mathrm{min}, t+5\mathrm{min}]$, as shown in the lower half of Fig.~\ref{fig:case_queue}. For all five days, we can see the arrival rates are larger than the service rates from around 7:00 to 7:50 am, and queues are therefore formed. The maximum arrival rate is often more than 300 people/min, while the maximum service rate is only around 200 people/min. The queue lengths start to decrease after around 7:50 am and the queue dissipates at around 9:00 am.

\begin{table}[htbp]
    \centering\small
    \caption{Quantifying the out-of-station queueing from 7:00 to 9:00 am.}
      \begin{tabular}{lrrrrrr}
      \toprule
            & Monday & Tuesday & Wednesday & Thursday & Friday & Average \\
      \midrule
      Total number of passengers & 22740 & 23193 & 22964 & 23241 & 22524 & 22932 \\
      Maximum waiting time (min) & 15.1  & 15.0  & 15.3  & 13.5  & 11.1  & 14.0 \\
      Total waiting time (hour) & 3665  & 2984  & 3240  & 3006  & 2194  & 3018 \\
      Average out-of-station waiting time (min) & 9.7   & 7.7   & 8.5   & 7.8   & 5.8   & 7.9 \\
      Maximum queue length (people) & 3191  & 3188  & 3251  & 2893  & 2310  & 2967 \\
      Maximum arrival rate (people/min) & 295   & 384   & 319   & 302   & 280   & 316 \\
      Maximum service rate (people/min) & 212   & 220   & 219   & 220   & 215   & 217 \\
      The time with the longest queue & 7:48  & 7:43  & 7:44  & 7:52  & 7:49  & 7:47 \\
      \bottomrule
      \end{tabular}%
    \label{tab:result}%
\end{table}%

Tab.~\ref{tab:result} summarizes major indices for the out-of-station queueing. We can see the out-of-station waiting greatly impact passengers' travel. The maximum arrival rate is 50\% larger than the maximum service rate. When the queue has a maximum length, over three thousand passengers are waiting in the queue, and it takes around fifteen minutes for a passenger to enter the station. On average, every passenger waits eight minutes outside of the station in the morning peak. Considering the number of passengers, the total out-of-station waiting time exceeds three thousand hours per day at the TTY-N station; the queueing is a big waste of time and efficiency.

\section{Potential solutions for out-of-station queueing} \label{sec:remedies}

Queueing outside of metro stations has a substantial negative impact on passenger travel experience. In this section, we discuss existing and potential solutions to this problem. The fundamental reason for the out-of-station waiting is the mismatch between demand and supply. The commuting demand is rooted in the urban structure and can hardly be changed. Many traffic congestion problems should be avoided in the initial urban planning stage. However, we can still manage the demand from a temporal aspect \citep{halvorsen2019demand}. For example, providing reduced-rate fares to off-peak trips can flatten the peak-hour demand (e.g., shift peak-hour trips to pre-peak and after-peak hours). The temporally differentiated fare scheme has been studied in many research \citep{yang2018managing, lu2020managing, li2018modeling, adnan2020examining}. A few real-world practices show that properly designed off-peak discounts can help reduce metro crowding \citep{halvorsen2016reducing, greene2018bart}. Based on the queueing analysis of Section~\ref{sec:case_queue}, a potential solution is to design a fare scheme for the TTY-N metro station to reduce the boarding demand from 7:00 to 8:00 am. Overall, using a temporally differentiated fare scheme is a potential solution, although the effect is hard to evaluate in advance.

Beijing metro has made a lot of efforts from the supply side. In fact, the minimum headway of most metro lines of Beijing has been reduced to less than two minutes to increase network capacity. Moreover, the original TTY-N metro station has been integrated into Tiantongyuan North Transportation Hub since October 13, 2019. The TTY-N Transportation Hub integrates metro line 5, coaches, buses, and a P+R parking lot. There is a large-scale waiting lobby in the hub, and passengers no longer need to queue in the open air, which is helpful under bad weather conditions. New transportation facilities are also under construction or planning. For example, the Beijing metro line 13A, which is expected to complete in 2023 \citep{NDRC2019}, can significantly relieve the commuting pressure of the Tiantongyuan area.

Reducing perceived waiting time can also improve the level of service. For example, improving the waiting environment by providing shelters \citep{fan2016waiting} and improving the thermal environment \citep{zhang2021outdoor} can significantly reduce the perceived waiting time. Besides, it has been shown that providing real-time information can reduce the anxiety for uncertainty and the perceived waiting time \citep{watkins2011my, brakewood2014experiment}. Therefore, providing real-time queueing information has the potential to improve transit services \citep{brakewood2015impact}.

Finally, using other transportation modes to share the metro's demand is also a solution. Normally, the bicycle is not an ideal substitution for the metro considering its short travel distance. However, there are always special cases. In 2019, a 6.5 km elevated bicycle-only path was built in Beijing to share the extremely high commuting demand between Huilongguan and Shangdi, where Huilongguan is another high-density residential area suffering from out-of-station queueing. It is reported that cycling between Huilongguan and Shangdi takes around 30 minutes, while it could take more than 40 minutes to commute by metro in the rush hour \citep{bike_only_2019}.

\section{Concluding Remarks}\label{sec:conclusion}

This paper proposes a data-driven method to estimate the waiting time outside of an oversaturated metro station due to flow control measures. To the best of our knowledge, this paper presents the first quantitative study to measure passengers' out-of-station waiting time. By combining smart card data from the metro and bus system, we use transfer passengers as a proxy to quantify the queueing time outside of a metro station. A probabilistic approach by Gaussian Process regression is developed to infer the out-of-station waiting time for all passengers. Besides, we propose to analyze the queueing process by a queueing diagram. In the TTY-N metro station case study, results show our method is robust to the noise in data and provides a reliable estimation for out-of-station waiting time. We find the out-of-station waiting can be a big burden---more than 15 minutes waiting time---for passengers in oversaturated metro stations. Our results could help transit agencies better understand service performance. In addition, the accurate estimation of out-of-station waiting can also be used to evaluate user utility and social cost, which could be further used to support decision making, such as design better flow-control strategies.

A limitation of this work is the lack of validation by a field survey. Because we can only access the smart card data of 2015, but the boarding demand at the TTY-N station has drastically changed since the completion of the TTY-N Transportation Hub and the outbreak of COVID-19. Nevertheless, the GP regression produces reasonable confident intervals, and we believe that our estimation should be a solid reference.

There are several directions for future research. First, the performance of metro systems can be re-evaluated by taking the out-of-station waiting into account, especially for megacities like Beijing. Our case study shows that passengers at Tiantongyuan North suffer while downstream passengers benefit from the flow-control measures. A critical research question is to balance the trade-off and design optimized flow-control strategies based on passenger flow assignment when demand exceeds network capacity. Second, because waiting in an open-air is more vulnerable to extreme weather, it is important to quantify the disutility of out-of-station waiting time \citep{tirachini2016valuation, zhang2021outdoor}. Furthermore, how the waiting time affects mode choice is also worthy of investigation \citep{sun2012rail}. Finally, an interesting and important research direction is to develop time- and station-dependent transit fare schemes to flatten peak hour demand and thus reduce the mismatch between demand and supply \citep{yang2018managing, lu2020managing, li2018modeling, adnan2020examining}.

\section*{Acknowledgement}
This research is supported by the Fonds de Recherche du Qu\'{e}bec - Soci\'{e}t\'{e} et Culture (FRQSC) under the NSFC-FRQSC Research Program on Smart Cities and Big Data and the Canada Foundation for Innovation (CFI) John R. Evans Leaders Fund.

\bibliographystyle{elsarticle-harv}
\bibliography{metro}

\begin{thebibliography}{46}
\expandafter\ifx\csname natexlab\endcsname\relax\def\natexlab#1{#1}\fi
\providecommand{\url}[1]{\texttt{#1}}
\providecommand{\href}[2]{#2}
\providecommand{\path}[1]{#1}
\providecommand{\DOIprefix}{doi:}
\providecommand{\ArXivprefix}{arXiv:}
\providecommand{\URLprefix}{URL: }
\providecommand{\Pubmedprefix}{pmid:}
\providecommand{\doi}[1]{\href{http://dx.doi.org/#1}{\path{#1}}}
\providecommand{\Pubmed}[1]{\href{pmid:#1}{\path{#1}}}
\providecommand{\bibinfo}[2]{#2}
\ifx\xfnm\relax \def\xfnm[#1]{\unskip,\space#1}\fi
%Type = Article
\bibitem[{Adnan et~al.(2020)Adnan, Biran, Baburajan, Basak and
  Ben-Akiva}]{adnan2020examining}
\bibinfo{author}{Adnan, M.}, \bibinfo{author}{Biran, B.h.N.},
  \bibinfo{author}{Baburajan, V.}, \bibinfo{author}{Basak, K.},
  \bibinfo{author}{Ben-Akiva, M.}, \bibinfo{year}{2020}.
\newblock \bibinfo{title}{Examining impacts of time-based pricing strategies in
  public transportation: A study of singapore}.
\newblock \bibinfo{journal}{Transportation Research Part A: Policy and
  Practice} \bibinfo{volume}{140}, \bibinfo{pages}{127--141}.
%Type = Article
\bibitem[{Brakewood et~al.(2014)Brakewood, Barbeau and
  Watkins}]{brakewood2014experiment}
\bibinfo{author}{Brakewood, C.}, \bibinfo{author}{Barbeau, S.},
  \bibinfo{author}{Watkins, K.}, \bibinfo{year}{2014}.
\newblock \bibinfo{title}{An experiment evaluating the impacts of real-time
  transit information on bus riders in tampa, florida}.
\newblock \bibinfo{journal}{Transportation Research Part A: Policy and
  Practice} \bibinfo{volume}{69}, \bibinfo{pages}{409--422}.
%Type = Article
\bibitem[{Brakewood et~al.(2015)Brakewood, Macfarlane and
  Watkins}]{brakewood2015impact}
\bibinfo{author}{Brakewood, C.}, \bibinfo{author}{Macfarlane, G.S.},
  \bibinfo{author}{Watkins, K.}, \bibinfo{year}{2015}.
\newblock \bibinfo{title}{The impact of real-time information on bus ridership
  in new york city}.
\newblock \bibinfo{journal}{Transportation Research Part C: Emerging
  Technologies} \bibinfo{volume}{53}, \bibinfo{pages}{59--75}.
%Type = Misc
\bibitem[{{China Daily}(2021)}]{bike_only_2019}
\bibinfo{author}{{China Daily}}, \bibinfo{year}{2021}.
\newblock \bibinfo{title}{Bike-only roadway connects {Beijing} communities}.
\newblock \URLprefix
  \url{https://www.chinadaily.com.cn/a/201905/31/WS5cf0765ea3104842260beca3.html}.
  \bibinfo{note}{{Accessed May 19, 2021}}.
%Type = Article
\bibitem[{Fan et~al.(2016)Fan, Guthrie and Levinson}]{fan2016waiting}
\bibinfo{author}{Fan, Y.}, \bibinfo{author}{Guthrie, A.},
  \bibinfo{author}{Levinson, D.}, \bibinfo{year}{2016}.
\newblock \bibinfo{title}{Waiting time perceptions at transit stops and
  stations: Effects of basic amenities, gender, and security}.
\newblock \bibinfo{journal}{Transportation Research Part A: Policy and
  Practice} \bibinfo{volume}{88}, \bibinfo{pages}{251--264}.
%Type = Book
\bibitem[{Gelman et~al.(2013)Gelman, Carlin, Stern, Dunson, Vehtari and
  Rubin}]{gelman2013bayesian}
\bibinfo{author}{Gelman, A.}, \bibinfo{author}{Carlin, J.B.},
  \bibinfo{author}{Stern, H.S.}, \bibinfo{author}{Dunson, D.B.},
  \bibinfo{author}{Vehtari, A.}, \bibinfo{author}{Rubin, D.B.},
  \bibinfo{year}{2013}.
\newblock \bibinfo{title}{Bayesian data analysis}.
\newblock \bibinfo{publisher}{CRC press}.
%Type = Misc
\bibitem[{{GPy}(since 2012)}]{gpy2014}
\bibinfo{author}{{GPy}}, \bibinfo{year}{since 2012}.
\newblock \bibinfo{title}{{GPy}: A gaussian process framework in python}.
\newblock \bibinfo{howpublished}{\url{http://github.com/SheffieldML/GPy}}.
%Type = Article
\bibitem[{Greene-Roesel et~al.(2018)Greene-Roesel, Castiglione, Guiriba and
  Bradley}]{greene2018bart}
\bibinfo{author}{Greene-Roesel, R.}, \bibinfo{author}{Castiglione, J.},
  \bibinfo{author}{Guiriba, C.}, \bibinfo{author}{Bradley, M.},
  \bibinfo{year}{2018}.
\newblock \bibinfo{title}{Bart perks: using incentives to manage transit
  demand}.
\newblock \bibinfo{journal}{Transportation Research Record}
  \bibinfo{volume}{2672}, \bibinfo{pages}{557--565}.
%Type = Article
\bibitem[{Halvorsen et~al.(2016)Halvorsen, Koutsopoulos, Lau, Au and
  Zhao}]{halvorsen2016reducing}
\bibinfo{author}{Halvorsen, A.}, \bibinfo{author}{Koutsopoulos, H.N.},
  \bibinfo{author}{Lau, S.}, \bibinfo{author}{Au, T.}, \bibinfo{author}{Zhao,
  J.}, \bibinfo{year}{2016}.
\newblock \bibinfo{title}{Reducing subway crowding: analysis of an off-peak
  discount experiment in hong kong}.
\newblock \bibinfo{journal}{Transportation Research Record}
  \bibinfo{volume}{2544}, \bibinfo{pages}{38--46}.
%Type = Article
\bibitem[{Halvorsen et~al.(2019)Halvorsen, Koutsopoulos, Ma and
  Zhao}]{halvorsen2019demand}
\bibinfo{author}{Halvorsen, A.}, \bibinfo{author}{Koutsopoulos, H.N.},
  \bibinfo{author}{Ma, Z.}, \bibinfo{author}{Zhao, J.}, \bibinfo{year}{2019}.
\newblock \bibinfo{title}{Demand management of congested public transport
  systems: a conceptual framework and application using smart card data}.
\newblock \bibinfo{journal}{Transportation} , \bibinfo{pages}{1--29}.
%Type = Article
\bibitem[{Jin et~al.(2014)Jin, Tang, Sun and Lee}]{jin2014enhancing}
\bibinfo{author}{Jin, J.G.}, \bibinfo{author}{Tang, L.C.},
  \bibinfo{author}{Sun, L.}, \bibinfo{author}{Lee, D.H.}, \bibinfo{year}{2014}.
\newblock \bibinfo{title}{Enhancing metro network resilience via localized
  integration with bus services}.
\newblock \bibinfo{journal}{Transportation Research Part E: Logistics and
  Transportation Review} \bibinfo{volume}{63}, \bibinfo{pages}{17--30}.
%Type = Article
\bibitem[{Jin et~al.(2015)Jin, Teo and Odoni}]{jin2015optimizing}
\bibinfo{author}{Jin, J.G.}, \bibinfo{author}{Teo, K.M.},
  \bibinfo{author}{Odoni, A.R.}, \bibinfo{year}{2015}.
\newblock \bibinfo{title}{Optimizing bus bridging services in response to
  disruptions of urban transit rail networks}.
\newblock \bibinfo{journal}{Transportation Science} \bibinfo{volume}{50},
  \bibinfo{pages}{790--804}.
%Type = Article
\bibitem[{Jyl{\"a}nki et~al.(2011)Jyl{\"a}nki, Vanhatalo and
  Vehtari}]{jylanki2011robust}
\bibinfo{author}{Jyl{\"a}nki, P.}, \bibinfo{author}{Vanhatalo, J.},
  \bibinfo{author}{Vehtari, A.}, \bibinfo{year}{2011}.
\newblock \bibinfo{title}{Robust gaussian process regression with a student-t
  likelihood.}
\newblock \bibinfo{journal}{Journal of Machine Learning Research}
  \bibinfo{volume}{12}.
%Type = Article
\bibitem[{Kang et~al.(2015)Kang, Wu, Sun, Zhu and Gao}]{kang2015case}
\bibinfo{author}{Kang, L.}, \bibinfo{author}{Wu, J.}, \bibinfo{author}{Sun,
  H.}, \bibinfo{author}{Zhu, X.}, \bibinfo{author}{Gao, Z.},
  \bibinfo{year}{2015}.
\newblock \bibinfo{title}{A case study on the coordination of last trains for
  the beijing subway network}.
\newblock \bibinfo{journal}{Transportation Research Part B: Methodological}
  \bibinfo{volume}{72}, \bibinfo{pages}{112--127}.
%Type = Article
\bibitem[{Krishnakumari et~al.(2020)Krishnakumari, Cats and van
  Lint}]{krishnakumari2020estimation}
\bibinfo{author}{Krishnakumari, P.}, \bibinfo{author}{Cats, O.},
  \bibinfo{author}{van Lint, H.}, \bibinfo{year}{2020}.
\newblock \bibinfo{title}{Estimation of metro network passenger delay from
  individual trajectories}.
\newblock \bibinfo{journal}{Transportation Research Part C: Emerging
  Technologies} \bibinfo{volume}{117}, \bibinfo{pages}{102704}.
%Type = Article
\bibitem[{Kusakabe et~al.(2010)Kusakabe, Iryo and
  Asakura}]{kusakabe2010estimation}
\bibinfo{author}{Kusakabe, T.}, \bibinfo{author}{Iryo, T.},
  \bibinfo{author}{Asakura, Y.}, \bibinfo{year}{2010}.
\newblock \bibinfo{title}{Estimation method for railway passengers’ train
  choice behavior with smart card transaction data}.
\newblock \bibinfo{journal}{Transportation} \bibinfo{volume}{37},
  \bibinfo{pages}{731--749}.
%Type = Article
\bibitem[{Li et~al.(2018)Li, Li, Xu, Liu and Ran}]{li2018modeling}
\bibinfo{author}{Li, H.}, \bibinfo{author}{Li, X.}, \bibinfo{author}{Xu, X.},
  \bibinfo{author}{Liu, J.}, \bibinfo{author}{Ran, B.}, \bibinfo{year}{2018}.
\newblock \bibinfo{title}{Modeling departure time choice of metro passengers
  with a smart corrected mixed logit model-a case study in beijing}.
\newblock \bibinfo{journal}{Transport Policy} \bibinfo{volume}{69},
  \bibinfo{pages}{106--121}.
%Type = Misc
\bibitem[{Li(2015)}]{beijing_queue_2019}
\bibinfo{author}{Li, P.}, \bibinfo{year}{2015}.
\newblock \bibinfo{title}{Six more subway stations to limit rush hour crowds}.
\newblock \URLprefix
  \url{https://www.thebeijinger.com/blog/2015/01/12/five-more-subway-stations-limit-rush-hour-crowds}.
  \bibinfo{note}{{The Beijinger, Accessed May 25, 2021}}.
%Type = Misc
\bibitem[{Lin(2021)}]{Shenzhen_queue_2021}
\bibinfo{author}{Lin, W.}, \bibinfo{year}{2021}.
\newblock \bibinfo{title}{After experiencing the morning peak of nanshan
  science and technology park, my mentality collapsed}.
\newblock \URLprefix
  \url{https://min.news/en/news/06f5d575e0557c7b52fa3ad7bbdabb85.html}.
  \bibinfo{note}{{MINEWS, Accessed May 25, 2021}}.
%Type = Article
\bibitem[{Lu et~al.(2020)Lu, Zhang, Long and Li}]{lu2020managing}
\bibinfo{author}{Lu, X.S.}, \bibinfo{author}{Zhang, X.}, \bibinfo{author}{Long,
  J.}, \bibinfo{author}{Li, Y.}, \bibinfo{year}{2020}.
\newblock \bibinfo{title}{Managing rail transit peak-hour congestion with step
  fare schemes}.
\newblock \bibinfo{journal}{Transportmetrica A: Transport Science}
  \bibinfo{volume}{16}, \bibinfo{pages}{1490--1511}.
%Type = Article
\bibitem[{Ma et~al.(2019)Ma, Koutsopoulos, Chen and Wilson}]{ma2019estimation}
\bibinfo{author}{Ma, Z.}, \bibinfo{author}{Koutsopoulos, H.N.},
  \bibinfo{author}{Chen, Y.}, \bibinfo{author}{Wilson, N.H.},
  \bibinfo{year}{2019}.
\newblock \bibinfo{title}{Estimation of denied boarding in urban rail systems:
  alternative formulations and comparative analysis}.
\newblock \bibinfo{journal}{Transportation Research Record}
  \bibinfo{volume}{2673}, \bibinfo{pages}{771--778}.
%Type = Article
\bibitem[{Mo et~al.(2020)Mo, Ma, Koutsopoulos and Zhao}]{mo2020capacity}
\bibinfo{author}{Mo, B.}, \bibinfo{author}{Ma, Z.},
  \bibinfo{author}{Koutsopoulos, H.N.}, \bibinfo{author}{Zhao, J.},
  \bibinfo{year}{2020}.
\newblock \bibinfo{title}{Capacity-constrained network performance model for
  urban rail systems}.
\newblock \bibinfo{journal}{Transportation Research Record}
  \bibinfo{volume}{2674}, \bibinfo{pages}{59--69}.
%Type = Misc
\bibitem[{{National Development and Reform Comission}(2019)}]{NDRC2019}
\bibinfo{author}{{National Development and Reform Comission}},
  \bibinfo{year}{2019}.
\newblock \bibinfo{title}{Reply on adjusting the planning scheme for the second
  phase of beijing urban rail transit ({No}. 1904)}.
\newblock \URLprefix
  \url{https://www.ndrc.gov.cn/xxgk/zcfb/tz/201912/t20191223_1216003.html}.
  \bibinfo{note}{{Accessed May 19, 2021}}.
%Type = Article
\bibitem[{Neal(1997)}]{neal1997monte}
\bibinfo{author}{Neal, R.M.}, \bibinfo{year}{1997}.
\newblock \bibinfo{title}{Monte carlo implementation of gaussian process models
  for bayesian regression and classification}.
\newblock \bibinfo{journal}{Technical Report 9702, Dept. of statistics and
  Dept. of Computer Science, University of Toronto} .
%Type = Article
\bibitem[{Niu and Zhou(2013)}]{niu2013optimizing}
\bibinfo{author}{Niu, H.}, \bibinfo{author}{Zhou, X.}, \bibinfo{year}{2013}.
\newblock \bibinfo{title}{Optimizing urban rail timetable under time-dependent
  demand and oversaturated conditions}.
\newblock \bibinfo{journal}{Transportation Research Part C: Emerging
  Technologies} \bibinfo{volume}{36}, \bibinfo{pages}{212--230}.
%Type = Article
\bibitem[{Pelletier et~al.(2011)Pelletier, Tr{\'e}panier and
  Morency}]{pelletier2011smart}
\bibinfo{author}{Pelletier, M.P.}, \bibinfo{author}{Tr{\'e}panier, M.},
  \bibinfo{author}{Morency, C.}, \bibinfo{year}{2011}.
\newblock \bibinfo{title}{Smart card data use in public transit: A literature
  review}.
\newblock \bibinfo{journal}{Transportation Research Part C: Emerging
  Technologies} \bibinfo{volume}{19}, \bibinfo{pages}{557--568}.
%Type = Article
\bibitem[{Qu et~al.(2020)Qu, Xu and Chien}]{qu2020estimating}
\bibinfo{author}{Qu, H.}, \bibinfo{author}{Xu, X.}, \bibinfo{author}{Chien,
  S.}, \bibinfo{year}{2020}.
\newblock \bibinfo{title}{Estimating wait time and passenger load in a
  saturated metro network: A data-driven approach}.
\newblock \bibinfo{journal}{Journal of Advanced Transportation}
  \bibinfo{volume}{2020}.
%Type = Article
\bibitem[{Shi et~al.(2019)Shi, Yang, Yang, Zhou and Gao}]{shi2019cooperative}
\bibinfo{author}{Shi, J.}, \bibinfo{author}{Yang, L.}, \bibinfo{author}{Yang,
  J.}, \bibinfo{author}{Zhou, F.}, \bibinfo{author}{Gao, Z.},
  \bibinfo{year}{2019}.
\newblock \bibinfo{title}{Cooperative passenger flow control in an
  oversaturated metro network with operational risk thresholds}.
\newblock \bibinfo{journal}{Transportation Research Part C: Emerging
  Technologies} \bibinfo{volume}{107}, \bibinfo{pages}{301--336}.
%Type = Inproceedings
\bibitem[{Sun et~al.(2015a)Sun, Jin, Lee and Axhausen}]{sun2015characterizing}
\bibinfo{author}{Sun, L.}, \bibinfo{author}{Jin, J.G.}, \bibinfo{author}{Lee,
  D.H.}, \bibinfo{author}{Axhausen, K.W.}, \bibinfo{year}{2015}a.
\newblock \bibinfo{title}{Characterizing multimodal transfer time using smart
  card data: the effect of time, passenger age, crowdedness, and collective
  pressure}, in: \bibinfo{booktitle}{Proceedings of Transportation Research
  Board 94th Annual Meeting}, \bibinfo{address}{Washington DC, United States}.
%Type = Article
\bibitem[{Sun et~al.(2014)Sun, Jin, Lee, Axhausen and Erath}]{sun2014demand}
\bibinfo{author}{Sun, L.}, \bibinfo{author}{Jin, J.G.}, \bibinfo{author}{Lee,
  D.H.}, \bibinfo{author}{Axhausen, K.W.}, \bibinfo{author}{Erath, A.},
  \bibinfo{year}{2014}.
\newblock \bibinfo{title}{Demand-driven timetable design for metro services}.
\newblock \bibinfo{journal}{Transportation Research Part C: Emerging
  Technologies} \bibinfo{volume}{46}, \bibinfo{pages}{284--299}.
%Type = Inproceedings
\bibitem[{Sun et~al.(2012)Sun, Lee, Erath and Huang}]{sun2012using}
\bibinfo{author}{Sun, L.}, \bibinfo{author}{Lee, D.H.}, \bibinfo{author}{Erath,
  A.}, \bibinfo{author}{Huang, X.}, \bibinfo{year}{2012}.
\newblock \bibinfo{title}{Using smart card data to extract passenger's
  spatio-temporal density and train's trajectory of mrt system}, in:
  \bibinfo{booktitle}{Proceedings of the ACM SIGKDD international workshop on
  urban computing}, \bibinfo{organization}{ACM}. pp. \bibinfo{pages}{142--148}.
%Type = Article
\bibitem[{Sun et~al.(2015b)Sun, Lu, Jin, Lee and Axhausen}]{sun2015integrated}
\bibinfo{author}{Sun, L.}, \bibinfo{author}{Lu, Y.}, \bibinfo{author}{Jin,
  J.G.}, \bibinfo{author}{Lee, D.H.}, \bibinfo{author}{Axhausen, K.W.},
  \bibinfo{year}{2015}b.
\newblock \bibinfo{title}{An integrated bayesian approach for passenger flow
  assignment in metro networks}.
\newblock \bibinfo{journal}{Transportation Research Part C: Emerging
  Technologies} \bibinfo{volume}{52}, \bibinfo{pages}{116--131}.
%Type = Article
\bibitem[{Sun and Xu(2012)}]{sun2012rail}
\bibinfo{author}{Sun, Y.}, \bibinfo{author}{Xu, R.}, \bibinfo{year}{2012}.
\newblock \bibinfo{title}{Rail transit travel time reliability and estimation
  of passenger route choice behavior: Analysis using automatic fare collection
  data}.
\newblock \bibinfo{journal}{Transportation Research Record}
  \bibinfo{volume}{2275}, \bibinfo{pages}{58--67}.
%Type = Article
\bibitem[{Tirachini et~al.(2016)Tirachini, Sun, Erath and
  Chakirov}]{tirachini2016valuation}
\bibinfo{author}{Tirachini, A.}, \bibinfo{author}{Sun, L.},
  \bibinfo{author}{Erath, A.}, \bibinfo{author}{Chakirov, A.},
  \bibinfo{year}{2016}.
\newblock \bibinfo{title}{Valuation of sitting and standing in metro trains
  using revealed preferences}.
\newblock \bibinfo{journal}{Transport Policy} \bibinfo{volume}{47},
  \bibinfo{pages}{94--104}.
%Type = Article
\bibitem[{Vanhatalo et~al.(2009)Vanhatalo, Jyl{\"a}nki and
  Vehtari}]{vanhatalo2009gaussian}
\bibinfo{author}{Vanhatalo, J.}, \bibinfo{author}{Jyl{\"a}nki, P.},
  \bibinfo{author}{Vehtari, A.}, \bibinfo{year}{2009}.
\newblock \bibinfo{title}{Gaussian process regression with student-t
  likelihood}.
\newblock \bibinfo{journal}{Advances in neural information processing systems}
  \bibinfo{volume}{22}, \bibinfo{pages}{1910--1918}.
%Type = Article
\bibitem[{Wardman(2004)}]{wardman2004public}
\bibinfo{author}{Wardman, M.}, \bibinfo{year}{2004}.
\newblock \bibinfo{title}{Public transport values of time}.
\newblock \bibinfo{journal}{Transport Policy} \bibinfo{volume}{11},
  \bibinfo{pages}{363--377}.
%Type = Article
\bibitem[{Watkins et~al.(2011)Watkins, Ferris, Borning, Rutherford and
  Layton}]{watkins2011my}
\bibinfo{author}{Watkins, K.E.}, \bibinfo{author}{Ferris, B.},
  \bibinfo{author}{Borning, A.}, \bibinfo{author}{Rutherford, G.S.},
  \bibinfo{author}{Layton, D.}, \bibinfo{year}{2011}.
\newblock \bibinfo{title}{Where is my bus? impact of mobile real-time
  information on the perceived and actual wait time of transit riders}.
\newblock \bibinfo{journal}{Transportation Research Part A: Policy and
  Practice} \bibinfo{volume}{45}, \bibinfo{pages}{839--848}.
%Type = Misc
\bibitem[{{Wikipedia contributors}(2019a)}]{beijingwiki}
\bibinfo{author}{{Wikipedia contributors}}, \bibinfo{year}{2019}a.
\newblock \bibinfo{title}{Beijing subway}.
\newblock \URLprefix \url{https://en.wikipedia.org/wiki/Beijing_Subway}.
  \bibinfo{note}{{Accessed July 18, 2019}}.
%Type = Misc
\bibitem[{{Wikipedia contributors}(2019b)}]{tiantongyuanwiki}
\bibinfo{author}{{Wikipedia contributors}}, \bibinfo{year}{2019}b.
\newblock \bibinfo{title}{Tiantongyuan}.
\newblock \URLprefix \url{https://en.wikipedia.org/wiki/Tiantongyuan}.
  \bibinfo{note}{{Accessed July 18, 2019}}.
%Type = Book
\bibitem[{Williams and Rasmussen(2006)}]{williams2006gaussian}
\bibinfo{author}{Williams, C.K.}, \bibinfo{author}{Rasmussen, C.E.},
  \bibinfo{year}{2006}.
\newblock \bibinfo{title}{Gaussian Processes for Machine Learning}.
\newblock \bibinfo{publisher}{The MIT Press}.
%Type = Article
\bibitem[{Xu et~al.(2019)Xu, Li, Liu, Ran and Qin}]{xu2019passenger}
\bibinfo{author}{Xu, X.}, \bibinfo{author}{Li, H.}, \bibinfo{author}{Liu, J.},
  \bibinfo{author}{Ran, B.}, \bibinfo{author}{Qin, L.}, \bibinfo{year}{2019}.
\newblock \bibinfo{title}{Passenger flow control with multi-station
  coordination in subway networks: algorithm development and real-world case
  study}.
\newblock \bibinfo{journal}{Transportmetrica B: Transport Dynamics}
  \bibinfo{volume}{7}, \bibinfo{pages}{446--472}.
%Type = Article
\bibitem[{Yang and Tang(2018)}]{yang2018managing}
\bibinfo{author}{Yang, H.}, \bibinfo{author}{Tang, Y.}, \bibinfo{year}{2018}.
\newblock \bibinfo{title}{Managing rail transit peak-hour congestion with a
  fare-reward scheme}.
\newblock \bibinfo{journal}{Transportation Research Part B: Methodological}
  \bibinfo{volume}{110}, \bibinfo{pages}{122--136}.
%Type = Article
\bibitem[{Yin et~al.(2016)Yin, Tang, Yang, Gao and Ran}]{yin2016energy}
\bibinfo{author}{Yin, J.}, \bibinfo{author}{Tang, T.}, \bibinfo{author}{Yang,
  L.}, \bibinfo{author}{Gao, Z.}, \bibinfo{author}{Ran, B.},
  \bibinfo{year}{2016}.
\newblock \bibinfo{title}{Energy-efficient metro train rescheduling with
  uncertain time-variant passenger demands: An approximate dynamic programming
  approach}.
\newblock \bibinfo{journal}{Transportation Research Part B: Methodological}
  \bibinfo{volume}{91}, \bibinfo{pages}{178--210}.
%Type = Article
\bibitem[{Zhang et~al.(2021)Zhang, Xu, Meng and Meng}]{zhang2021outdoor}
\bibinfo{author}{Zhang, M.}, \bibinfo{author}{Xu, C.}, \bibinfo{author}{Meng,
  L.}, \bibinfo{author}{Meng, X.}, \bibinfo{year}{2021}.
\newblock \bibinfo{title}{Outdoor comfort level improvement in the traffic
  waiting areas by using a mist spray system: An experiment and questionnaire
  study}.
\newblock \bibinfo{journal}{Sustainable Cities and Society} ,
  \bibinfo{pages}{102973}.
%Type = Article
\bibitem[{Zhao et~al.(2016)Zhao, Zhang, Tu, Xu, Shen, Tian, Li and
  Li}]{zhao2016estimation}
\bibinfo{author}{Zhao, J.}, \bibinfo{author}{Zhang, F.}, \bibinfo{author}{Tu,
  L.}, \bibinfo{author}{Xu, C.}, \bibinfo{author}{Shen, D.},
  \bibinfo{author}{Tian, C.}, \bibinfo{author}{Li, X.Y.}, \bibinfo{author}{Li,
  Z.}, \bibinfo{year}{2016}.
\newblock \bibinfo{title}{Estimation of passenger route choice pattern using
  smart card data for complex metro systems}.
\newblock \bibinfo{journal}{IEEE Transactions on Intelligent Transportation
  Systems} \bibinfo{volume}{18}, \bibinfo{pages}{790--801}.
%Type = Article
\bibitem[{Zhu et~al.(2018)Zhu, Koutsopoulos and Wilson}]{zhu2018inferring}
\bibinfo{author}{Zhu, Y.}, \bibinfo{author}{Koutsopoulos, H.N.},
  \bibinfo{author}{Wilson, N.H.}, \bibinfo{year}{2018}.
\newblock \bibinfo{title}{Inferring left behind passengers in congested metro
  systems from automated data}.
\newblock \bibinfo{journal}{Transportation Research Part C: Emerging
  Technologies} \bibinfo{volume}{94}, \bibinfo{pages}{323--337}.

\end{thebibliography}

\end{document}